\documentstyle[11pt,moriond-modified,epsfig]{article}

\newcommand{\met}{\mbox{$\not\!\!E_T$}}

\def\be{\begin{equation}}
\def\ee{\end{equation}}
\def\bea{\begin{eqnarray}}
\def\eea{\end{eqnarray}}

\begin{document}
\vspace*{4cm}
\title{RENCONTRES {\sc de} MORIOND '98 -- QCD AND HADRONIC INTERACTIONS\\
  EXPERIMENTAL SUMMARY}

\author{ MARK STROVINK}

\address{Department of Physics and Lawrence Berkeley National Laboratory, University of California\\
Berkeley, California 94720, USA}

\maketitle\abstracts{
I discuss new experimental results reported at this conference that bear directly on three previously reported anomalies: the excess of high $Q^2$ events at HERA; the excess of $W + 1$ jet events, relative to $W$'s without jets, in D\O\ data; and the excess of high $E_T$ inclusive jets observed by CDF.  The new results all point in the direction of reducing the experimental significance of these excesses.
}

\section{Introduction}
Near the close of this meeting it was my pleasure to provide its experimental summary.  I attempted to convey a few of the main points from 41 of the presentations, ranging from heavy ion interactions to top quark decay.  Shortly thereafter, I placed on the Web~\cite{web} one or two scanned summary transparencies for each of the 60 talks for which such material was supplied or prepared.  This information will remain accessible for several years.  However, in neither of these efforts was I able to cover every experimental presentation; for that I apologize.

Because of its volume, that Web material provides a more complete as well as a more timely record than I would be able to supply in a contribution to these Proceedings.  So I resolved here to focus on only a few experimental topics.  Where possible, in addition to reporting the facts, I aim to discuss a bit of the recent history and perhaps to add some criticism.  As for the broad directions charted by all of the superb work reported at this meeting, I defer to the theoretical summary~\cite{ellis} by Keith Ellis.

In his summary~\cite{collins} of this conference one year ago, the first three experimental topics mentioned by John Collins were the excess of high $Q^2$ events at HERA, the excess of high $E_T$ inclusive jets in CDF data, and the excess of $W + 1$ jet events, relative to $W$'s without jets, reported by D\O.  At this meeting, important new results were presented that bear directly on these topics, to which I now turn.

\section{High $Q^2$ $ep$ Scattering at HERA}

\subsection{HERA 1994-96 data}\label{subsec:96hera}

At this meeting one year ago, a topic of much discussion was the excess of events in neutral current data at high $Q^2$ reported in early 1997 by the H1~\cite{h196} and ZEUS~\cite{zeus96} collaborations.  The H1 excess appeared most prominently in a 25 GeV/c$^2$ bin of electron-quark invariant mass $M = \sqrt{sx_{\rm Bj}}$ centered at $M = 200$ GeV/c$^2$, with $Q^2/M^2 \equiv y > 0.4$ and $\sqrt{s}=300$ GeV.  Seven events were observed, while 0.95$\pm$0.18 were expected from neutral current processes [$P(\ge$7) = 0.026\%].  In the study of an ensemble of Monte Carlo experiments, fewer than 1\% yielded a less probable excess for any bin in the range $80 < M < 250$ GeV/c$^2$.  The ZEUS excess was localized in the region $y > 0.25$ and $x_{\rm Bj} > 0.55$ ($M > 223$ GeV/c$^2$), where 4 events were observed and 0.71$\pm$0.06 were expected (probability 0.6\%).  Two of the ZEUS events exceeded $Q^2$=35000 GeV/c$^2$, compared to 0.145$\pm$0.013 expected.  In the study of an ensemble of Monte Carlo experiments, fewer than 6\% yielded a less probable excess for any $Q^2$ threshold.

In a previously unexplored region, when one is faced with two similar but apparently independent anomalies such as these, thoroughly and responsibly reported, one must pursue vigorously the chance that they signal new physics, regardless of other possible prejudices {\it e.g.}~about the impregnability of the Standard Model, or about difficulties in measuring tails of distributions.  Both of these HERA groups did just that, along with many other experimenters and theorists worldwide.

\subsection{Remarks on HERA event kinematics}\label{subsec:96kin}

Nominally, the H1 and ZEUS $eq$ mass regions described above ($187.5 < M < 212.5$ and $M > 223$ GeV/c$^2$, respectively) are disjoint.  However, the experimental quantities measured by H1 and ZEUS were not identical; the kinematic variables quoted by H1 were based on electron energy and angle, while the ZEUS kinematics were based on jet and electron angles only.  Taking into account this difference as well as the effects of resolution and calibration uncertainty, Bassler and Bernardi~\cite{bassler} nevertheless were able to conclude later in 1997 that the two excesses indeed were inconsistent with arising from a resonant enhancement at a single $M$.

These differences in measurement methods were possible because, except for radiative effects, the events are twice constrained.  I think that the highest $Q^2$ events at HERA are precious enough to warrant performing a maximum likelihood fit simultaneously to all the measured energies and angles available for each event.  A general maximization, not requiring the parameter errors to be gaussian, could incorporate approximately the radiative effects as well.  If successful, such fits would yield not only the optimum kinematic resolution, but also a maximum likelihood whose distribution would be useful, in comparison to Monte Carlo, for assessing the extent to which the measured parameters of these rare events are internally consistent.

\begin{table}[b]
\caption{
Expected and observed preliminary totals of neutral and charged current $e^+p$ scattering events with $Q^2 > Q^2$(min), from full 1994-97 HERA datasets.
}
\label{tab:hera}
\vspace{0.4cm}
\centerline{\psfig{figure=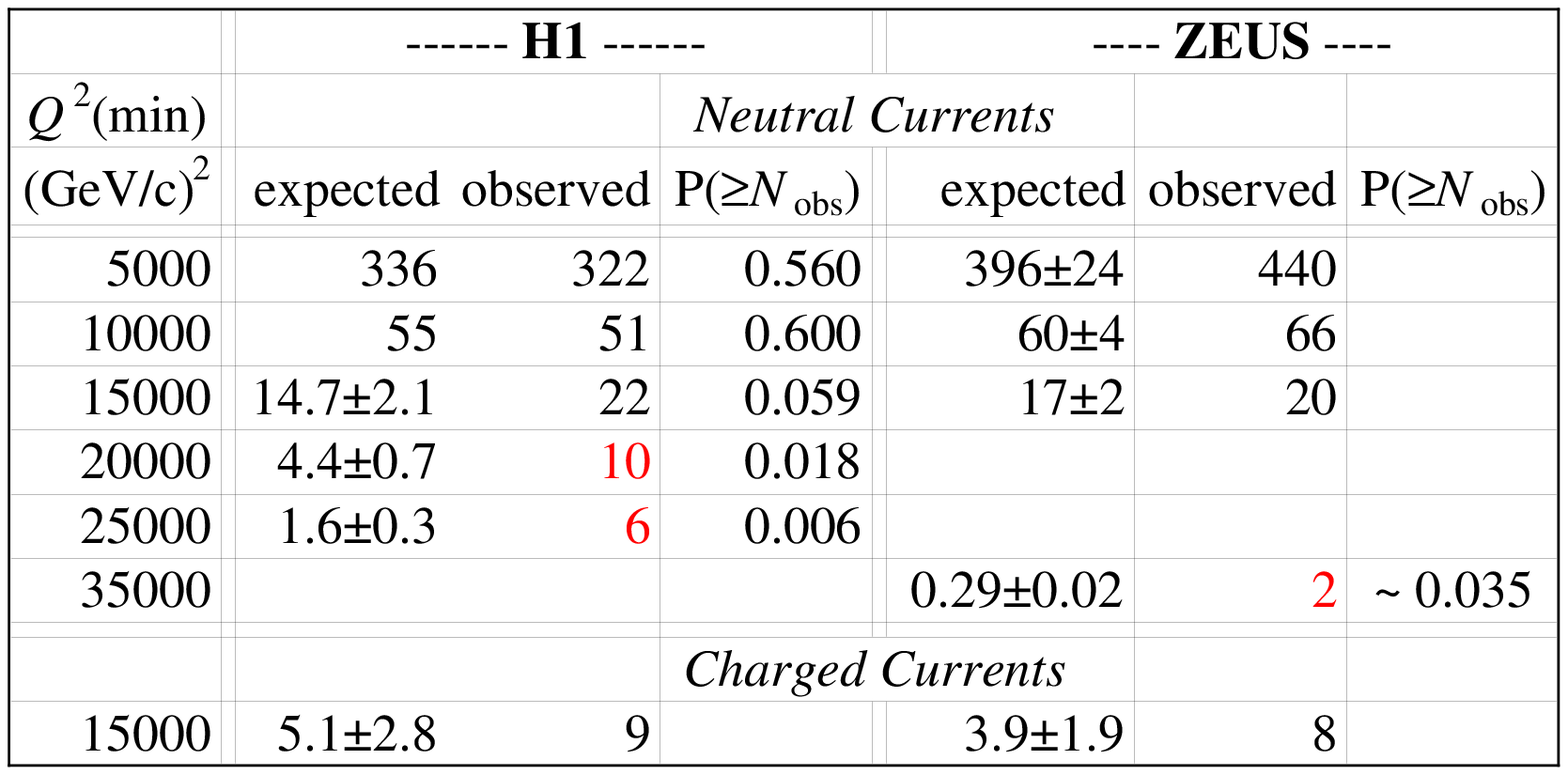,angle=0,width=4.2in,clip=t}}
\end{table}

\subsection{Leptoquark searches stimulated by the 1994-96 HERA data}\label{subsec:96lepto}

Meanwhile, provoked especially by the suggestion of a $\approx$200 GeV/c$^2$ peak in the H1 data, searches for first generation leptoquarks were intensified.  Particularly restrictive were those at the Tevatron, where pairs of leptoquarks can be produced in the $s$ channel by quark-antiquark annihilation, as are top quark pairs, regardless of the strength of the unknown electron-quark coupling.  (This coupling need only be large enough to cause the leptoquark to decay well within the detector.)  The remaining parameters are the leptoquark spin (0 or 1) and $\beta$, the leptoquark branching ratio to $eq$ as opposed to $\nu q$; theoretically~\cite{hewitt} $\beta$ is strongly favored to be unity, with $\beta = {1 \over 2}$ also a possibility.  By mid 1997, D\O~\cite{d0slq10} and CDF~\cite{cdfslq10} had submitted for publication 95\% C.L.~lower mass limits of 225 and 213 GeV/c$^2$, respectively, for $\beta =1$ scalar leptoquarks.  These were soon followed by D\O~\cite{d0slq05} and preliminary CDF~\cite{cdfslq05} limits of 204 and 180 GeV/c$^2$, respectively, for $\beta = 0.5$, and by a limit contour from D\O~\cite{d0slq05} for the full range $0 < \beta < 1$.  Most recently, D\O\ and CDF have combined their likelihoods for $\beta=1$ to obtain~\cite{landsberg} at 95\% confidence a lower mass limit for first generation scalar leptoquarks of 242 GeV/c$^2$.

While these limits on the masses of scalar leptoquarks were being set, it was generally appreciated that the corresponding limits on vector leptoquarks should be more stringent because their production cross section tends to be higher.  At this conference Boehnlein~\cite{d0vlq} presented new D\O\ mass limit contours {\it vs.}~$\beta$ for first generation leptoquarks with three possible vector couplings, one of which minimizes the production rate.  Even in that worst case, a 200 GeV/c$^2$ vector leptoquark is ruled out at 95\% C.L.~for $eq$ branching ratio $\beta > 0.14$.

 \subsection{HERA 1997 data}\label{97hera}

At this meeting, preliminary H1 and ZEUS results were presented by Glazov~\cite{h197} and Nagano~\cite{zeus97}, respectively, for the full 1994-97 datasets.  The total integrated luminosities of $\approx$14 and $\approx$20 pb$^{-1}$, respectively, that were available at the end of 1996 have increased to $\approx$37 and $\approx$47 pb$^{-1}$; calibrations have been refined and the acceptances improved.  For H1, above the same threshold ($y > 0.4$) and within the same 25 GeV/c$^2$ window centered at $M = 200$ GeV/c$^2$ where the largest excess was observed in the 1994-96 data, only one additional event was collected in 1997, with $\approx$2.1$\pm$0.5 events expected from Standard Model processes.  Therefore the new H1 data do not confirm the earlier indication of a clustering of excess events around $M = 200$ GeV/c$^2$.  Likewise, in the original ZEUS signal region $y > 0.25$ and $x_{\rm Bj} > 0.55$, the most recently available $x$-$y$ scatter plot~\cite{williams} from ZEUS for the entire 1994-97 data set shows no increase in the number of events originally reported on the basis of the 1994-96 data alone.

Keeping in mind that low-statistics samples indeed are expected to fluctuate from one subset to another, we follow the experimenters and attempt to characterize whatever residual excesses remain in the full H1 and ZEUS datasets.  In light of the negative results for the 1997 data in the particular signal regions just described, we use simple $Q^2$ thresholds to describe any remaining excesses.  These are summarized in Table~\ref{tab:hera}.  The largest fluctuation tabulated there is for H1 neutral current data with $Q^2 > 25000$ (GeV/c)$^2$, where 6 events are observed with 1.6$\pm$0.3 expected (probability 0.6\%).  (Presumably, in an ensemble of Monte Carlo experiments, the fraction of experiments exhibiting a fluctuation of probability $<$0.6\% above {\sl any} $Q^2$ threshold would be larger than 0.6\%.)  Smaller fluctuations also are observed for ZEUS neutral current data with $Q^2 > 35000$ (GeV/c)$^2$, and for both H1 and ZEUS charged current samples with $Q^2 > 15000$ (GeV/c)$^2$.  The significances of the charged current excesses are limited, however, by systematic errors in the expectations.  It should be kept in mind that not all of the entries in Table~\ref{tab:hera} are statistically independent.

\begin{figure}[t]
\centerline{\psfig{figure=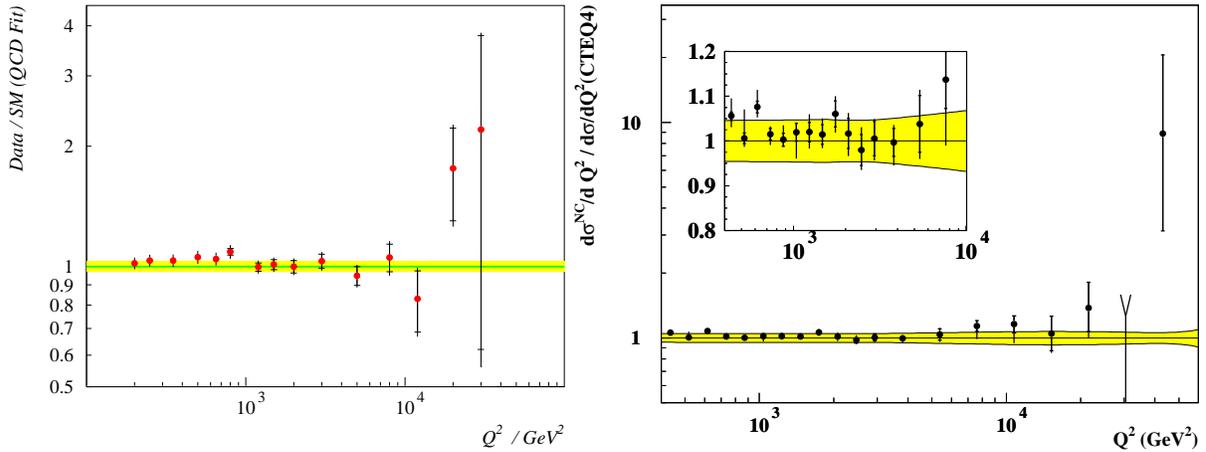,angle=0,width=6.25in,clip=t}}
\caption{
Preliminary ratio {\it vs.}~$Q^2$ of $d\sigma/dQ^2$ from data to that from a QCD NLO fit evolved from lower $Q^2$ data, for (left) H1 and (right) ZEUS full 1994-97 datasets.  Inner errors are statistical and outer errors are statistical+systematic in quadrature.  For H1 the shaded band is the overall luminosity error, which does not contribute to the error bars; for ZEUS the shaded band shows the effect of varying the parton density function.
}
\label{fig:h1-ratio-vs-q2-cropped}
\end{figure}

The issue of residual excesses at high $Q^2$ in the full HERA datasets is perhaps best summarized by Fig.~\ref{fig:h1-ratio-vs-q2-cropped}, which shows for each experiment the ratio {\it vs.}~$Q^2$ of $d\sigma/dQ^2$ for neutral current data to that for a QCD NLO fit evolved from lower $Q^2$ data.  Over this $Q^2$ range the cross section falls by seven decades, but the ratio plotted is nearly flat.  As in Table~\ref{tab:hera}, excesses are visible that may not be significant.  The effect of including the H1 data in the H1 NLO QCD fit is to {\sl reduce} $d\sigma/dQ^2$ at high $x_{\rm Bj}$ and $Q^2$.  Also, as demonstrated by a complementary plot, the effects expected in the neutral current channel from $\gamma Z$ interference have become visible.  

\section{$W$ Production at the Tevatron: $p_T(W)$ and Extra Jets}

\subsection{Transverse momenta of $W$ bosons produced by ${\bar p}p$ collisions}\label{subsec:ptw}

Two potentially related deviations from expectations for $W$ production at the Tevatron were reported at conferences in the first half of 1997 as the result of preliminary analyses by D\O.  Of the two, the less significant deviation~\cite{melanson} involved the shape of the $p_T$ spectrum of the $W$.  In this analysis of 1992-93 data, $p_T(W)$ was measured by forming the vector sum of all the energy deposits in the calorimeter, excluding the $W$ decay electron.  For $p_T(W) > 60$ GeV/c, the data lay above the ${\cal O}(\alpha_s^2)$ prediction by a factor $\approx 1.77 \pm 0.24$ (stat) $\pm 0.21$ (syst), or about 2.4$\sigma$.  (I supply this factor only approximately because it is inferred from a plot~\cite{melanson}, assuming the systematic errors to be fully correlated at each $p_T(W)$; numerical results were not presented for this preliminary analysis.)  The systematic error available at that time included the effects of uncertainties only in jet energy scale and resolution.

\begin{figure}[t]
\centerline{\hbox{\psfig{figure=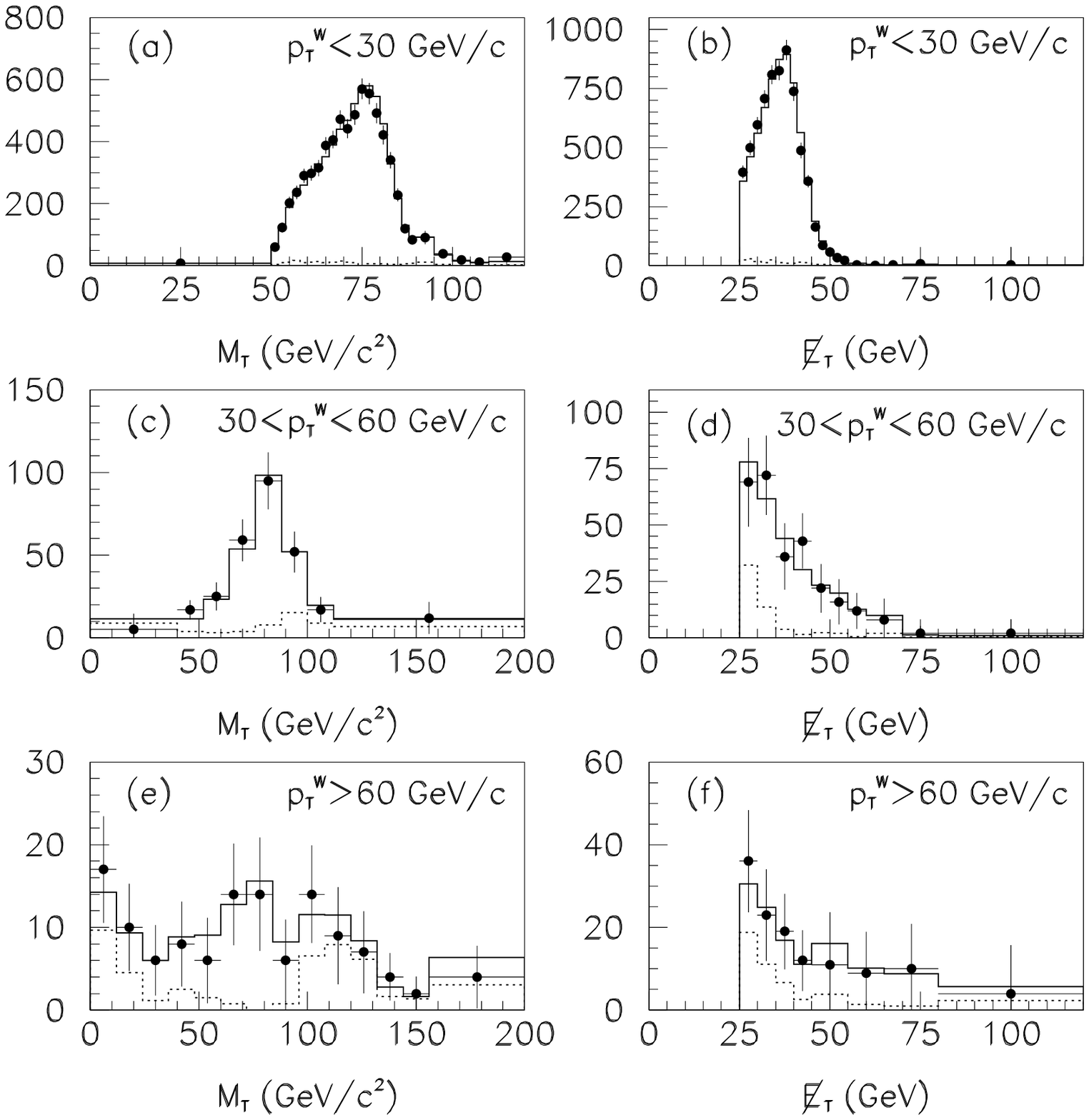,angle=0,height=3in,clip=t} \hglue 0.15in \psfig{figure=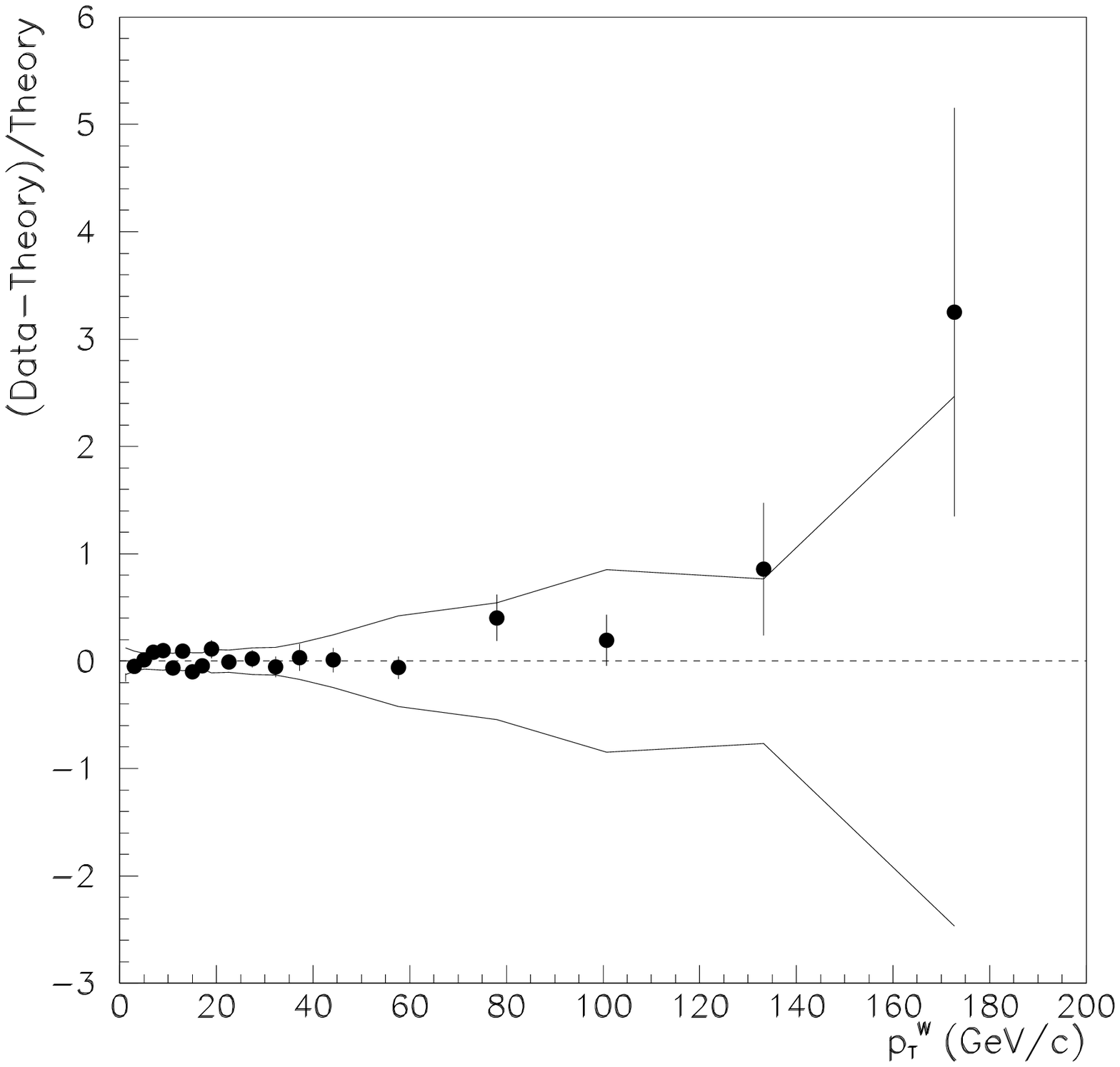,angle=0,height=3in,clip=t}}}
\caption{
(Left) Distribution in $e\nu$ transverse mass $M_T$ and in missing $E_T$ for production of $W$ bosons in three bins of $p_T(W)$ by 1.8 TeV $p {\bar p}$ collisions at D\O.  The points are data, the dashed histogram is estimated background from events with misidentified electrons, and the solid histogram is the sum of background and expectation from best fit $W \rightarrow e \nu$ signal.  (Right) Ratio {\it vs.}~$p_T(W)$ of background subtracted $W \rightarrow e \nu$ signal to NLO {\sc dyrad} calculation, normalized to the signal. 
}
\label{fig:gerber-fig}
\end{figure}

At this conference Gerber~\cite{gerber-talk} presented the final analysis of this dataset, which has since been published~\cite{gerber-prl}.  In the journey from a preliminary to a final measurement, much effort was devoted to understanding the background from multijets with misidentified electrons, which in this analysis turns out to be large (38\%$\pm$22\%) at high $p_T(W)$ ($>60$ GeV/c).  Such backgrounds conventionally are estimated using an event sample to which the electron identification criteria have not yet been applied.  The efficiency for multijet background events to satisfy these criteria is measured first for a background-rich region (missing $E_T \equiv$ \met\ $< 15$ GeV in this case).  The same background efficiency is then assumed to hold for events in the signal region (\met\ $> 25$ GeV).  In their published analysis, D\O\ progressed beyond this simple assumption, studying and allowing for a possible variation of the background misidentification efficiency with \met.  As a cross-check, they demanded that the sum of best fit expected signal and estimated background contributions describe the observed shapes of distributions in transverse $e\nu$ mass $M_T$ of the $W$ boson, as well as in \met.  These distributions vary markedly with $p_T(W)$, becoming more diffuse at high $p_T(W)$.  As demonstrated on the left-hand side of Fig.~\ref{fig:gerber-fig}, the agreement between data and expectation is very good.

The published ratio of data to NLO prediction, displayed on the right-hand side of Fig.~\ref{fig:gerber-fig}, exhibits no significant deviation from unity for $p_T(W) > 60$ GeV; a careful analysis shows that any discrepancy there is less than ${1 \over 2} \sigma$ in the total error.  The main evolution of this result from its preliminary version is a moderate increase in the central background estimate at high $p_T(W)$, and in the uncertainty assigned to that estimate.  Having completed this analysis, the D\O\ experimenters are well positioned to attack their much larger 1995-96 $W$ boson sample, to which more powerful electron identification techniques can be applied, including the use of transition radiation detector information.  The prospects seem bright at the Tevatron for more stringent tests of NLO calculations at high $p_T(W)$. 
 
\subsection{Extra jets in $W$ boson production at the Tevatron}\label{subsec:r10}

\begin{figure}[t]
\centerline{\hbox{\psfig{figure=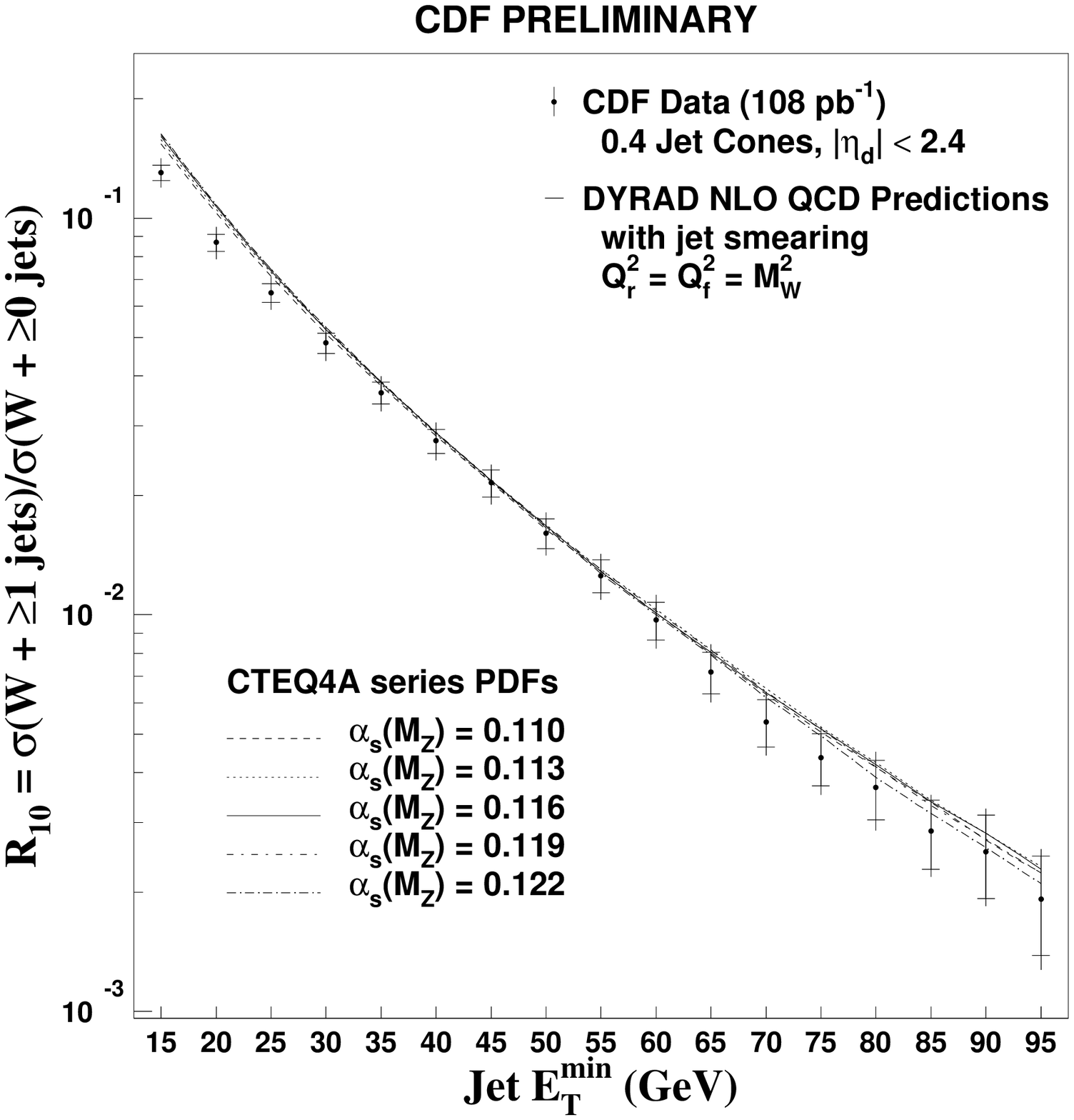,angle=0,height=2.8in,clip=t} \hglue 0.05in \psfig{figure=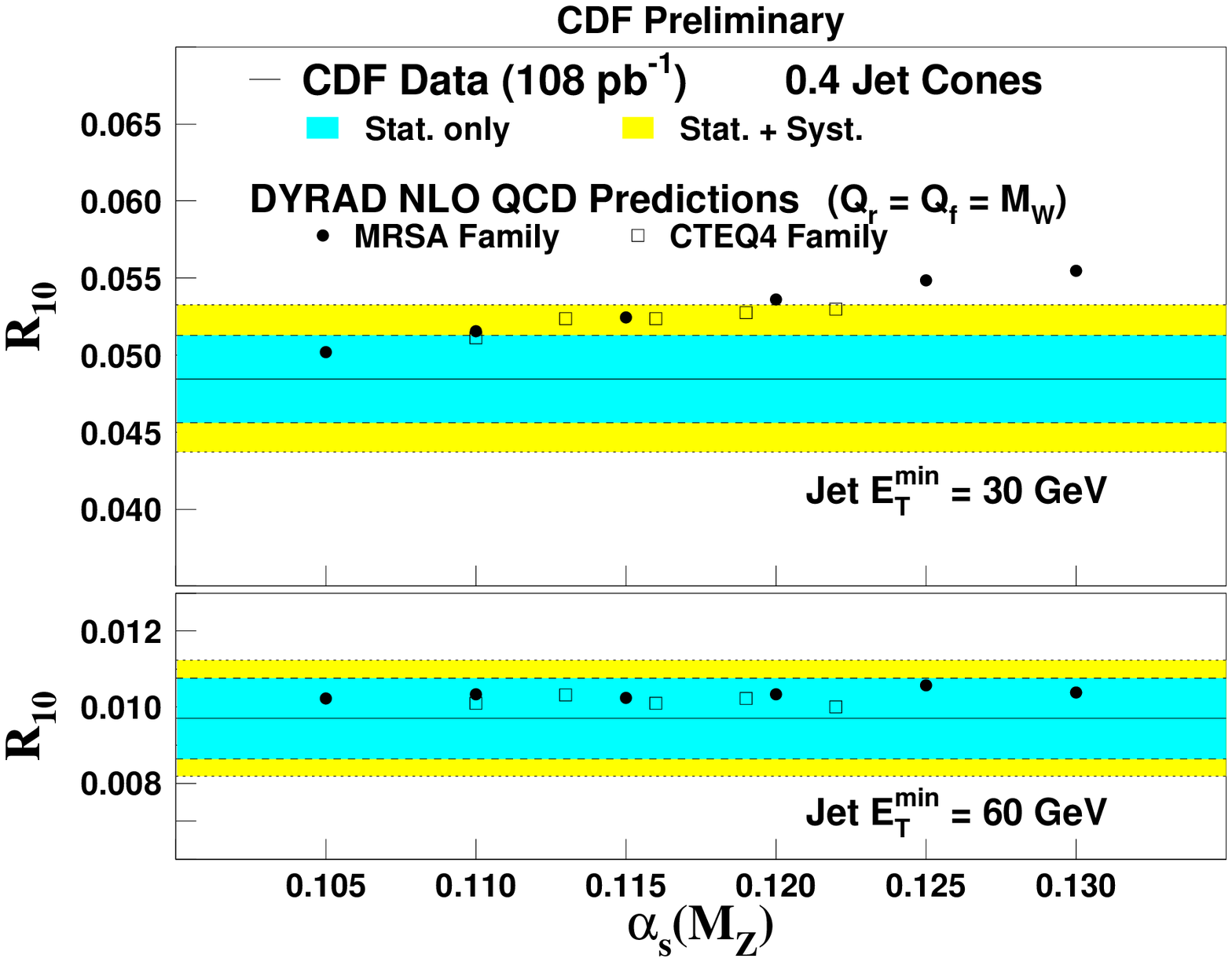,angle=0,height=2.8in,clip=t}}}
\caption{
(Left) ${\cal R}^{10}_{\rm incl}$ (see text) {\it vs.}~minimum jet $E_T$ from preliminary CDF analysis, compared with NLO {\sc dyrad} predictions for {\sc cteq4a} PDFs using five values of $\alpha_s$.  (Right) Dependence on $\alpha_s$ of calculated ${\cal R}^{10}_{\rm incl}$ (points) for two values of minimum jet $E_T$, compared to CDF preliminary observation (bands).
}
\label{fig:cdf-r10}
\end{figure}

The more significant of the two deviations from expectations for $W$ production, reported preliminarily by D\O\ in 1997, involved the quantity ${\cal R}^{10}_{\rm excl}$.  This is the ratio of the cross section for {\sl exclusive} production of $W$ plus exactly one jet to the cross section for $W$ plus zero jets, where the jets must exceed a transverse energy threshold $E_T^{\rm min}$.  At that time, for $E_T^{\rm min} = 25$ GeV, D\O\ plotted~\cite{joffe} a preliminary value ${\cal R}^{10}_{\rm excl} \approx 0.096 \pm  0.005$ (stat) $\pm 0.008$ (syst).  Because clustering the energy from a calorimeter into a jet and measuring its energy, with finite resolution, are specific to a particular experimental setup, this value of ${\cal R}^{10}_{\rm excl}$ is best compared to a NLO {\sc dyrad}~\cite{giele} calculation which includes these same experimental effects.  At that time, for $E_T^{\rm min} = 25$ GeV, D\O\ calculated and plotted values of ${\cal R}^{10}_{\rm excl}$ that ranged from $\approx$0.051 to 0.056, varying weakly with reasonable choices of $\alpha_s$ and parton density function.  For example, if one took the calculated ${\cal R}^{10}_{\rm excl}$ to be 0.053$\pm$0.002, it would lie 4.4$\sigma$ below the preliminary observation, which was a factor of $\approx$1.8 higher.  The discrepancy grew slightly with $E_T^{\rm min}$ out to $\approx$60 GeV/c, where the statistical precision ran out.

At this conference, presented on behalf of CDF by Gerber~\cite{gerber-talk} was a new preliminary determination of ${\cal R}^{10}_{\rm incl}$, the ratio of the cross section for {\sl inclusive} production of $W$ plus one or more jets to the total cross section for $W$ production.  These results are plotted {\it vs.}~$E_T^{\rm min}$ in the left-hand portion of Fig.~\ref{fig:cdf-r10}; for $E_T^{\rm min} = 25$ GeV, the plotted value of ${\cal R}^{10}_{\rm incl}$ is $\approx 0.065 \pm 0.004$ (stat) $\pm 0.005$ (syst).  Likewise using {\sc dyrad}, CDF plots a calculated value of ${\cal R}^{10}_{\rm incl}$ that ranges from $\approx$0.070 to 0.074 for $E_T^{\rm min} = 25$ GeV, in agreement with their observation.  This is in conflict with the 1997 D\O\ result; the ratio of CDF data to theory is $\approx$0.9 rather than $\approx$1.8.  The consistency of CDF data with theory persists out to $E_T^{\rm min} = 95$ GeV.  The right-hand part of Fig.~\ref{fig:cdf-r10} displays the weak variation of calculated ${\cal R}^{10}_{\rm incl}$ with $\alpha_s$ for two values of $E_T^{\rm min}$. 

Although ${\cal R}^{10}_{\rm excl}$ and ${\cal R}^{10}_{\rm incl}$ in principle are different quantities, with different theoretical uncertainties, they tend to track each other closely.  For example, in a ``Berends scaling'' model~\cite{berends} in which $\sigma(W + $($n$+1) jets$)/\sigma(W + n$ jets) is equal~\cite{cdf-wnjets} to a constant $\rho$, the difference between ${\cal R}^{10}_{\rm excl}$ and ${\cal R}^{10}_{\rm incl}$ is of order $\rho^3 < 0.001$ for all the cases considered here.  Different definitions of ${\cal R}^{10}$ cannot account for the conflict between D\O\ and CDF results.

Although both experiments use a cone algorithm in $\eta$-$\phi$ space for clustering their jets, the cone radii $R$ are different ($R = 0.7$ for D\O\ {\it vs.}~$R = 0.4$ for CDF).  Therefore one does not expect an exact correspondence between the two experimental values of ${\cal R}^{10}$.  Even after the jets are corrected for noise, underlying event, and multiple interactions, one would expect more jet energy from radiated gluons to be captured within the larger cone, raising the observed and calculated value of ${\cal R}^{10}$.  As long as the experimental conditions are different (in cone size, clustering algorithm, or jet resolution), the proper results with which to compare the two experiments are their respective ratios of data ${\cal R}^{10}$ to theory ${\cal R}^{10}$.

That having been said, it is curious nonetheless that D\O\ calculates a {\sl smaller} ${\cal R}^{10}$ ($\approx$0.053) for a three times larger {\sl larger} cone area than does CDF ($\approx$0.072).  The experimenters have checked~\cite{dittmann} that their results from {\sc dyrad} are consistent at the generated parton level.  But there is a qualitative difference in their procedures for modeling the experimental effects.  D\O\ smears the calorimeter energy using a gaussian with a width that is based on studies of jet $E_T$ balance using dijet data, and it models the effects of jet clustering by merging {\sc dyrad} partons that are closer than $(R_{\rm sep} = 1.3) \times (R = 0.7)$ in $\eta$-$\phi$ space.  CDF simulates the experimental effects using parametrized Monte Carlo $W$ (+ jets) events based on {\sc herwig}-fragmented partons generated at tree level by {\sc vecbos}.  These events are then reweighted to more closely resemble the data.  If their Monte Carlo is sufficiently well parametrized, CDF's approach should permit a more detailed correction for the effects of jet energy resolution, clustering, splitting and merging.  

Another difference in the two analyses concerns the relative correction for electron identification efficiency in 0-jet and 1-jet $W$ events.  Based on studies carried out using $Z$ (+ jets) events, D\O\ applies a significant correction~\cite{joffe} to ${\cal R}^{10}_{\rm excl}$ ($\times$1.2 for $E_T^{\rm min} = 25$ GeV) for this effect.  CDF, whose electron identification scheme is not the same as D\O's, applies a smaller correction based on Monte Carlo simulation.

With final results on ${\cal R}^{10}$ unavailable from either experiment at the time of this meeting, what can be concluded?  The foregoing discussion suggests two particular features that should be shared by a fully believable measurement of ${\cal R}^{10}$.  First, the analysis should be repeated for more than one cone radius $R$.  Over a reasonable range of $R$, the comparison of ${\cal R}^{10}$ for data to theory should lead to a conclusion that is qualitatively independent of $R$; otherwise one would question whether the energy correction, smearing, clustering, splitting or merging of jets is modeled reliably.  Second, many of the uncertainties in acceptance for jets which appear to lie above a threshold in $E_T$ involve the possibility of unmodeled energy leakage into the jet cone.  Such leakage can be exacerbated at high instantaneous luminosity ${\cal L}$.  Obviously the comparison of  ${\cal R}^{10}$ for data to theory should lead to conclusions that are  ${\cal L}$-independent.  Unfortunately, the precision with which this can be checked, by splitting the sample into several ${\cal L}$ bins, is coarser than the full-sample error on ${\cal R}^{10}$.  A complementary approach would be to identify variables that should be particularly sensitive to variations in ${\cal L}$, and to check that these variations are well modeled.  An example~\cite{heidi} is $p_T(W)$ (measured as was discussed in section~\ref{subsec:ptw}).  As Summers~\cite{summers} reminded us at this conference, at lowest order $p_T(W)$ should be distributed with a step at $E_T^{\rm min}$.  In addition to jet energy resolution and higher-order effects, multiple interactions and pileup at high ${\cal L}$ would be expected to smear that step, providing an opportunity for making this check.

\section{High $E_T$ Inclusive Jets at the Tevatron}

\subsection{CDF 1996 excess}\label{subsec:96cdf}

\begin{figure}[t]
\centerline{\hbox{\psfig{figure=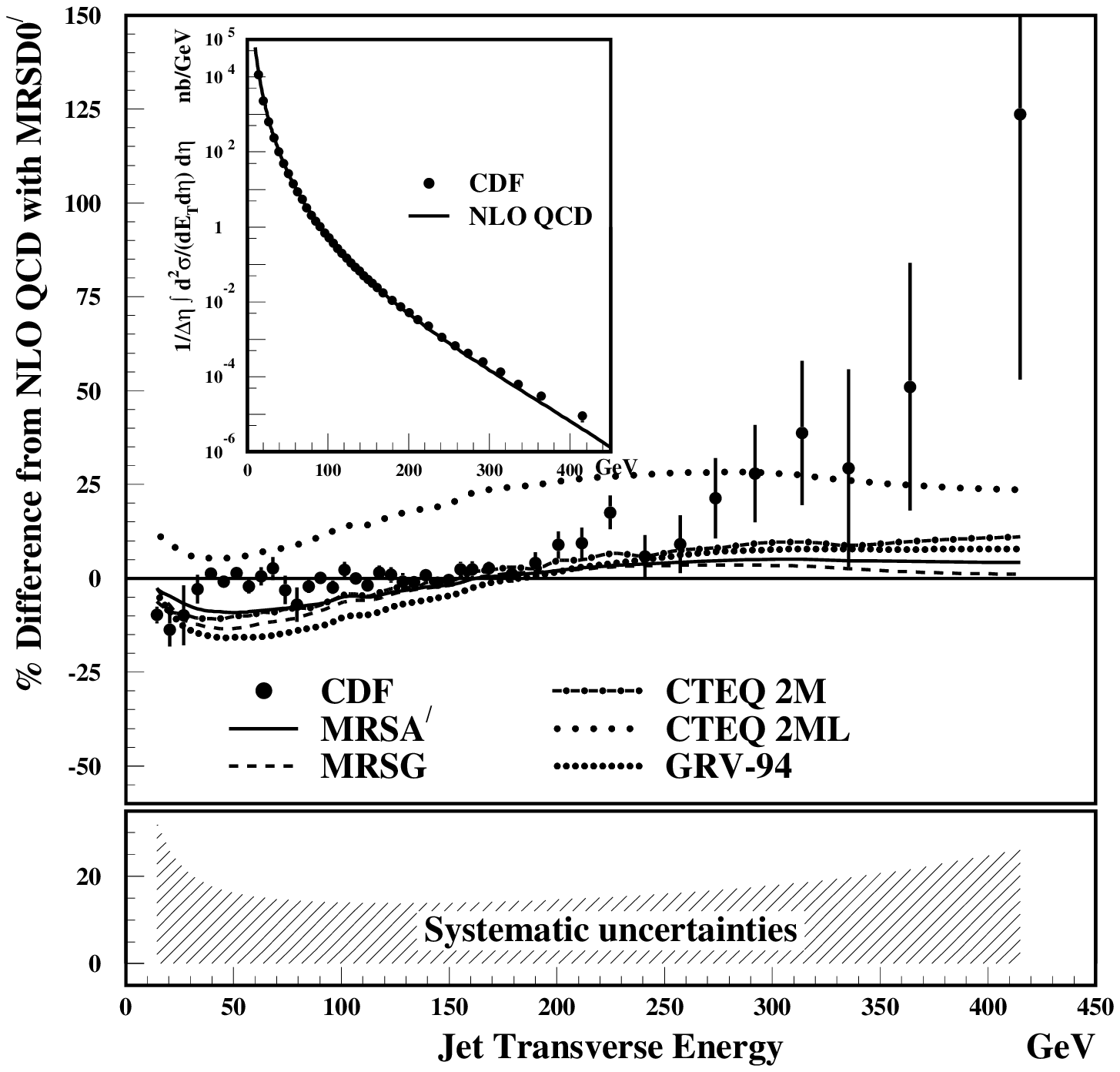,angle=0,height=3in,clip=t} \hglue 0.05in \psfig{figure=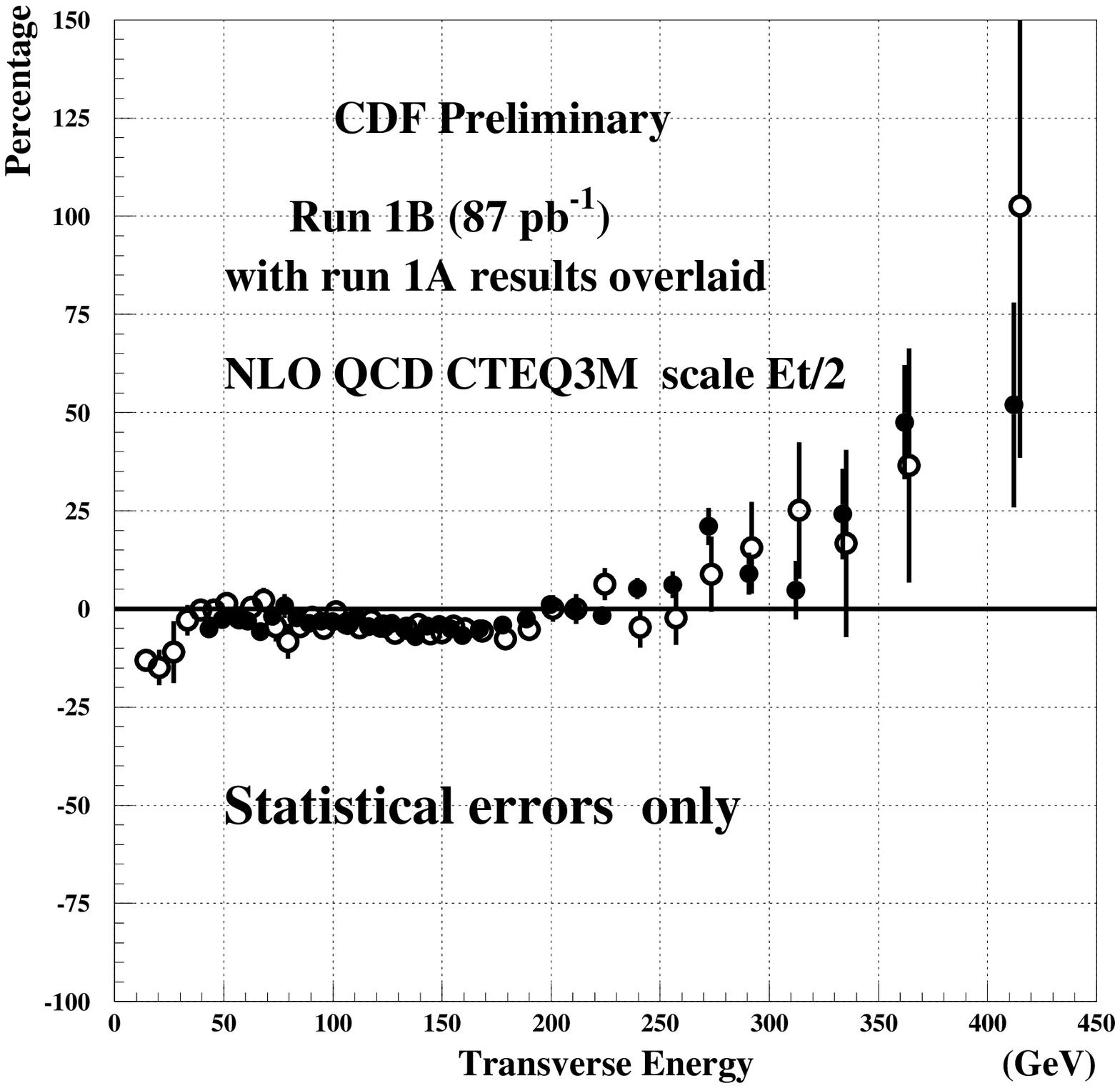,angle=0,height=3in,clip=t}}}
\caption{
(Left) Percentage difference (points) between the inclusive jet cross section from CDF 1992-93 data and a NLO {\sc eks} calculation using the {\sc mrsd0}$'$ PDF with $\mu = {1 \over 2}E_T^{\rm jet}$.  Errors are statistical; systematic uncertainties are shown by the shaded band below.  The lines show the variation in the calculation when other PDFs are substituted.  The inset displays {\it vs.}~$E_T$ the cross section differential in $E_T$ and $\eta$, averaged over $0.1 < |\eta| < 0.7$.  (Right) Same percentage for the same 1992-93 data (open points) and for the preliminary 1994-95 CDF data (filled points), relative to the same calculation using the {\sc cteq3m} PDF.
}
\label{fig:cdf-incl}
\end{figure}

As is well known, early in 1996 CDF published~\cite{cdf-incl1a} their observation of an excess of high $E_T$ inclusive jet production in $p{\bar p}$ collisions at $\sqrt{s} = 1.8$ TeV, relative to NLO calculations based on standard parton distribution functions (PDFs).  Their key plot, based on 1992-93 data, is reproduced on the left-hand side of Fig.~\ref{fig:cdf-incl}.  The excess is statistically significant; the authors state that no single source of systematic uncertainty can account for it.  Parameters are supplied for a ``standard curve'' that passes smoothly through the measured cross section points.  

Shortly thereafter, CDF reported~\cite{cdf-incl1b} a confirmatory preliminary result from their 1995-1996 sample (right-hand side of Fig.~\ref{fig:cdf-incl}), with improved statistical errors.  As one rough measure of the effect in both datasets, between $\approx$200 $< E_T <$ $\approx$400 GeV the ratio  {\it vs.}~$E_T$ of CDF data to expectation, based on the {\sc cteq3m} PDF, appears to acquire a slope of $\approx$2 TeV$^{-1}$, corresponding to a $\approx$40\% rise over this range.

The inclusive jet data reported~\cite{blazey} by D\O\ in 1996, and updated~\cite{elvira-old} in 1997, were consistent in shape with calculations based on the {\sc cteq3m} PDF; D\O\ did not confirm the CDF excess.  However, calibration of the calorimeter energy scale, which is the main systematic uncertainty in both measurements, had not yet been completed by D\O; in 1996 their systematic errors were larger than those of CDF.  The two groups' measurements were not considered to be in mutual conflict.

Phenomenologically, an economical way to accommodate the CDF excess is to perturb the PDFs (particularly the gluon PDF), which are determined mainly by deep inelastic lepton-nucleon scattering data.  Because it is difficult to parametrize the (correlated) systematic uncertainties in these input data, some groups who perform these PDF fits adjust by hand the statistical weights of individual inputs.  When, to obtain their {\sc cteq4hj} PDF,  the {\sc cteq} group~\cite{cteq} awarded to CDF data in the region of the excess an artificially large weight, they deemed the overall fit still to be of acceptable quality despite a rise~\cite{brock} of 23 units in its $\chi^2$.  Relative to {\sc cteq3m}, the inclusive jet cross section calculated on the basis of the {\sc cteq4hj} PDF increases~\cite{bertram} by $\approx$21\% between $E_T$ = 200 and 400 GeV, parametrizing about half of the CDF anomaly.

\subsection{Final D\O\ results on inclusive jets}

\begin{figure}[t]
\centerline{\hbox{\psfig{figure=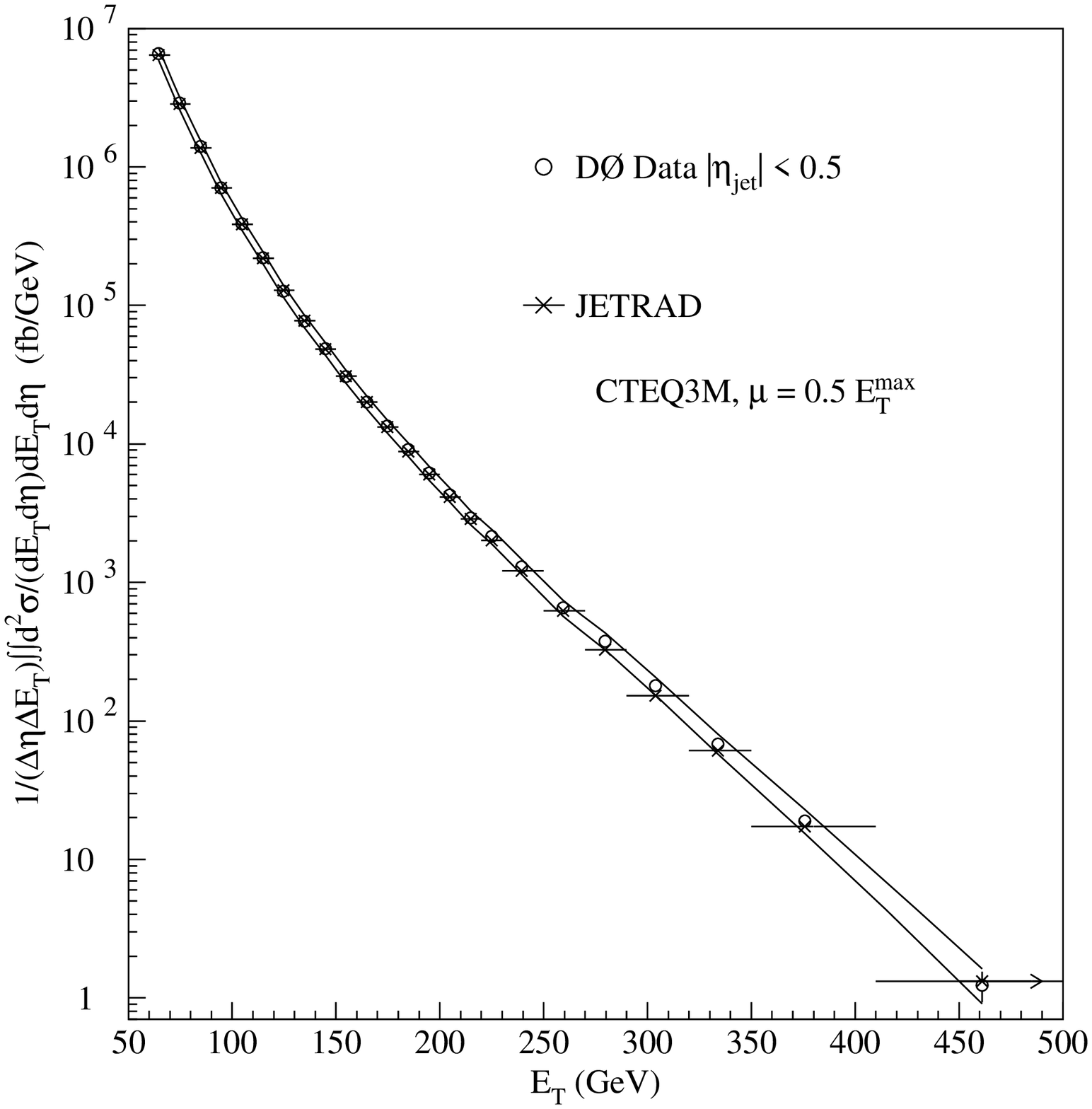,angle=0,height=3.07in,width=2.1in,clip=t} \hglue 0.05in \psfig{figure=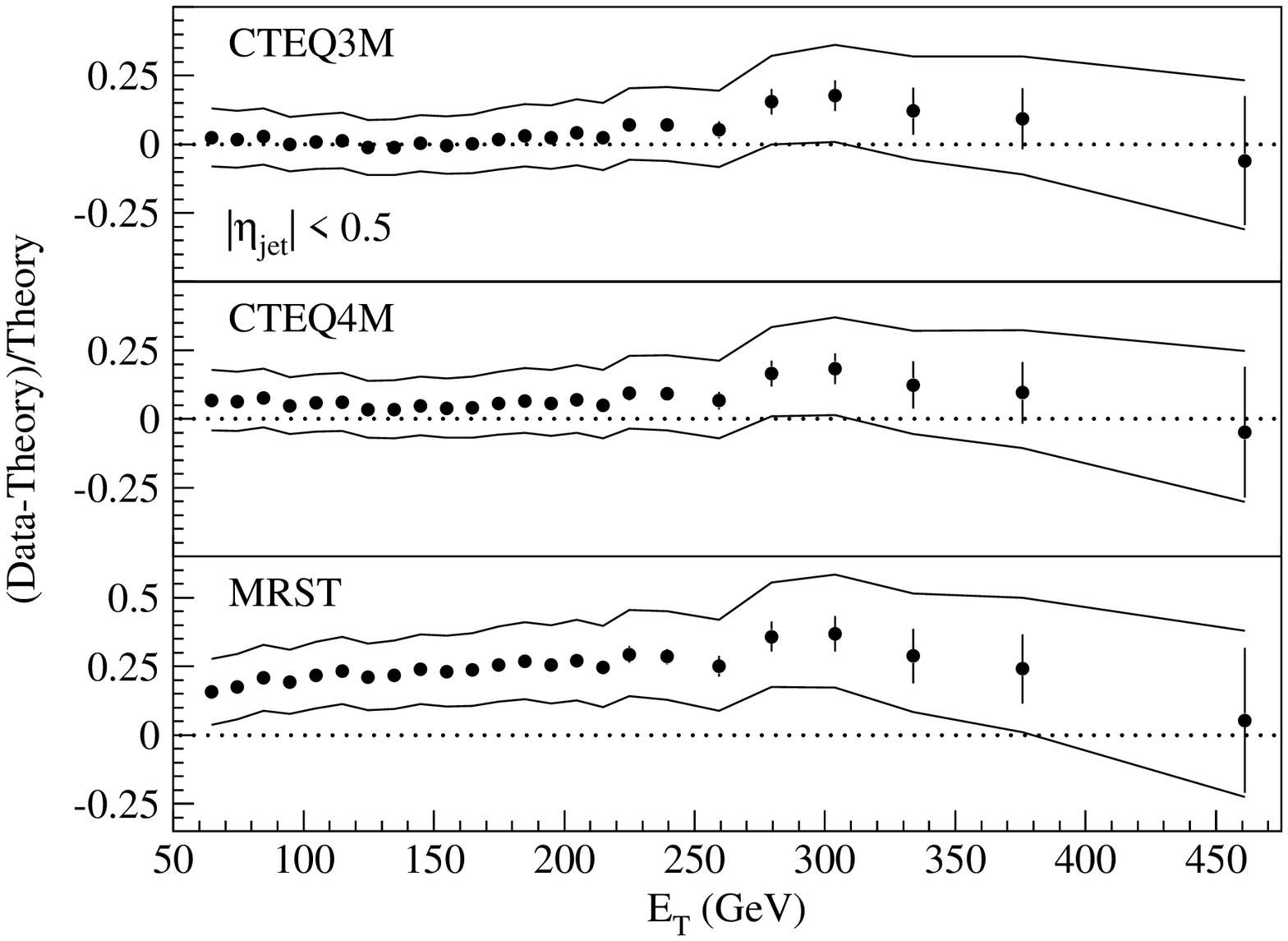,angle=0,height=3in,clip=t}}}
\caption{
(Left) Inclusive jet cross section (crosses) {\it vs.}~$E_T$, differential in $E_T$ and $\eta$, averaged over $|eta|<0.5$ and the indicated bins in $E_T$, from 1994-95 D\O\ data.  The lines show the $\pm 1\sigma$ systematic error excursions; the statistical errors are invisible on this scale.  The open circles are the results of a NLO {\sc jetrad} calculation using the {\sc cteq3m} PDF with $\mu = {1 \over 2}E_T^{\rm max}$.  (Right) Fractional deviation (points) of the same measured cross section from the same calculation, using the {\sc cteq3m}, {\sc cteq4m}, and {\sc mrst} PDFs.  The errors are statistical; the lines show the $\pm 1\sigma$ systematic uncertainties.
}
\label{fig:d0-incl}
\end{figure}

By the time of this meeting, D\O\ had completed~\cite{cafix} its calorimeter calibration, based on {\it in situ} transverse energy balance in $\gamma$+jet events.  For $R = 0.7$ jets with $|\eta|<0.5$, the jet energy at $E_T =$ 100 (300) GeV is corrected by a factor of $\approx$1.15 (1.13), which is determined with a fractional uncertainty of 1.5\% (1.8\%).  This uncertainty induces a 7.5\% (13\%) error on $d\sigma/dE_T$ for inclusive jets, which dominates the 10\% (15\%) total systematic error.  (In 1996, CDF's fractional systematic error in $d\sigma/dE_T$ was larger by about 0.03 at both energies.)  D\O's systematic uncertainty in $d\sigma/dE_T$ is parametrized by an error covariance matrix, indexed by the discrete values of $E_T$ at which $d\sigma/dE_T$ is measured.  This permits the calculation of a $\chi^2$ and an associated confidence interval for any functional fit to these points.

The final D\O\ inclusive jet cross section from 1995-96 data was presented at this conference by Elvira~\cite{elvira}.  It is exhibited in Fig.~\ref{fig:d0-incl}, both absolutely and as a ratio to a {\sc jetrad}~\cite{jetrad} calculation using three modern PDFs, for $|\eta|<0.5$ and a renormalization scale $\mu = 0.5 E_T^{\rm max}$, where $E_T^{\rm max}$ is the maximum jet $E_T$ in the event.  The $\chi^2$ probabilities are at least $\approx$50\% for the hypotheses that these calculations describe the D\O\ measurements.  They remain satisfactory if the {\sc eks} code~\cite{eks} is substituted~\cite{bertram} for {\sc jetrad}; if the range $0.1 < |\eta| < 0.7$ is substituted for the range $|\eta|<0.5$; or if $\mu$ is equated to $c{\cal E}_T$, with $c = {1 \over 4}, {1 \over 2}$, or 1, and ${\cal E}_T = E_T^{\rm max}$ or $E_T^{\rm jet}$.  Therefore, the D\O\ data do not require or suggest any deviation between the measured inclusive jet cross section and current NLO calculations based on modern standard PDFs.

\begin{figure}[t]
\centerline{\psfig{figure=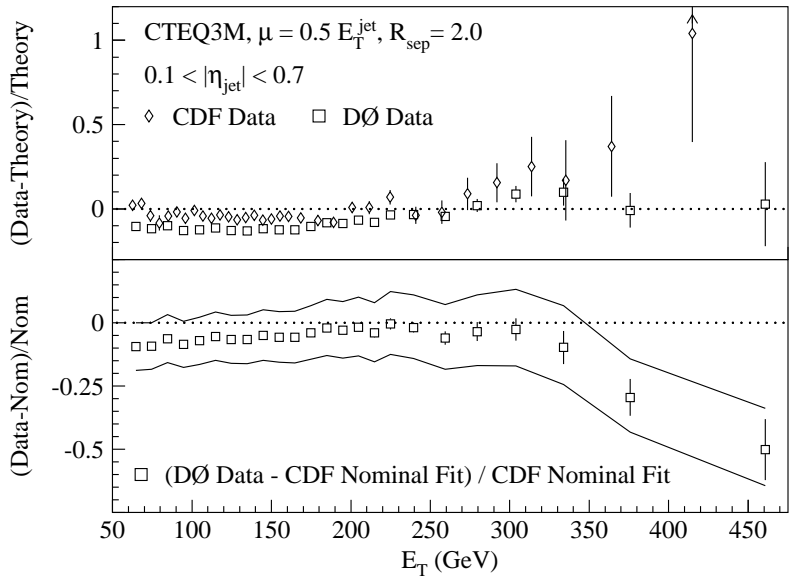,angle=0,height=3in,clip=t}}
\caption{
Comparison {\it vs.}~$E_T$ of inclusive jet cross sections, differential in $E_T$ and $\eta$ and averaged over $0.1 < |\eta| < 0.7$, from 1992-93 CDF (diamonds) and 1994-95 D\O\ (squares) data.  (Above) Fractional difference of either cross section from the same NLO {\sc eks} calculation using the parameters shown.  (Below) Fractional difference of the D\O\ cross section from the ``standard curve'' published by CDF to parametrize their measured cross section.  All errors are statistical; the lines reflect the $\pm 1\sigma$ systematic uncertainties in the D\O\ cross section.
}
\label{fig:cdf-d0-incl}
\end{figure}

Figure~\ref{fig:cdf-d0-incl} exhibits two types of comparison of the final results for 1994-95 D\O\ data to the published results for 1992-93 CDF data~\cite{apology}.  In the top panel, the two sets of data points are presented as a ratio to the same calculation, allowing a rigorous mutual comparison.  Only statistical errors are shown.  One gains the impression that, if reasonable systematic errors are taken into account, the two datasets still are not in mutual conflict.  This impression is reinforced by the construction~\cite{heidi-covariance} and use of an approximate error covariance matrix for the CDF points.

The bottom panel in Fig.~\ref{fig:cdf-d0-incl} shows the ratio of D\O\ data to CDF's aforementioned published  ``standard curve''.  Again, this is a rigorous comparison of D\O\ and CDF data, with the latter represented by the curve; theory plays no role in this ratio.  The $\chi^2$ probability for the hypothesis that the curve describes the D\O\ data is $\approx$$2 \times 10^{-5}$.  Therefore, the D\O\ data reject the CDF ``standard curve'' at $>4 \sigma$, and, by implication, the particular deviations from theory which that curve may represent.

\section{Closing Discussion}

I have attempted to describe a current, and, where possible, quantitative experimental status in three areas where possible deviations from QCD calculations have been noted in 1997 or 1996:  excess events at high $Q^2$ in HERA data, aspects of $W$ production in D\O\ data, and excess inclusive jets at high $E_T$ in CDF data.  In many cases not all of the anticipated information is in hand; it would be immensely exciting if one or more of these effects were to become well established and able successfully to challenge the Standard Model.  However, in these areas, the new experimental results that were reported to this conference {\sl all} point in the direction of reducing the experimental significance of these anomalies.

One is reminded of the 1972 article~\cite{pakvasa} {\it A Unified Explanation of the} $K^0_L \rightarrow \mu^+ \mu^-$ {\it Puzzle, CP Nonconservation, and the Excess Muon Anomaly}.  The fact that hardly anybody remembers two of these anomalies is a clue that they, in particular, have a unified explanation; unfortunately it is more pedestrian than the article's authors had hoped.  Among the most stimulating and enriching pursuits for the remaining four-dimensional theorists in our field, and even for the experimentalists, is to build models and ingenious physical explanations for unexpected observations in the data.  It is also productive, if not always as stimulating, to continue to scrutinize closely the basic experimental physics of those observations -- and to increase the statistics.   

\section*{Acknowledgments}

I am grateful to J.~Tran Thanh Van and the Organizing Committee for the opportunity to provide this summary, and for their splendid hospitality.  This work was supported by the U.S.~Department of Energy under Contract No.~DE-AC03-76F00098.


\begin{thebibliography}{99}
\bibitem{apology} Ideally the preliminary 1994-95 CDF data would also be included here.  However, the available plot was prepared for a D\O\ journal article, where it would not be appropriate for those data first to be published.
\end{thebibliography}
\end{document}